\documentclass{ws-procs9x6_mg}
\usepackage{epsfig}

\begin{document}

\title{
%\vspace*{-1cm}
%\rightline{CDF/PUB/CDF/PUBLIC/8304} ~~ \\ 
Diffractive and exclusive \\ measurements at CDF\footnote{\uppercase{P}resented at the 
``\uppercase{XIV} \uppercase{I}nternational \uppercase{W}orkshop
on \uppercase{D}eep \uppercase{I}nelastic \uppercase{S}cattering'' (\uppercase{DIS2006}), 
\uppercase{T}sukuba, \uppercase{J}apan, 20-24 \uppercase{A}pril, 2006}}

\author{Michele Gallinaro\footnote{\uppercase{R}epresenting the \uppercase{CDF} collaboration.}}

\address{Laboratory of Experimental High Energy Physics\\
The Rockefeller University\\
1230 York Avenue, New York, NY 10021, USA}

\maketitle

\abstracts{
Experimental results from the CDF experiment at the Tevatron in $p\bar{p}$ collisions at $\sqrt{s}$=1.96~TeV are presented on
the diffractive structure function at different values of the exchanged
momentum transfer squared in the range $0<Q^2<10,000$~GeV$^2$, on the four-momentum transfer $|t|$ distribution
in the region $0<|t|<1$~GeV$^2$ for both soft and hard diffractive events up to $Q^2\approx 4,500$~GeV$^2$, and on
the first experimental evidence of exclusive production in both dijet and diphoton events.
A novel technique to align the Roman Pot detectors is also presented.}

\section{Quantum Chromodynamics and diffraction}

Diffractive processes are characterized by a final state
in which a large region of rapidity is not filled with particles (``rapidity gap'')
and where the incident hadrons that survive are emitted at small angles with respect to the original beam direction.
The traditional ``pomeron'' can be defined within the framework of Quantum Chromodynamics (QCD)
and can be described as a composite entity of quarks and gluons\cite{dino}.
The goal of the CDF diffractive studies is two-fold:
(a) to obtain results which can help decipher the QCD nature of the Pomeron, 
such as the measurement of the diffractive structure function (DSF)
and $|t|$ distributions, and 
(b) to measure exclusive production rates (dijet, $\chi_c^0$, $\gamma\gamma$),
which could be used to establish benchmark calibrations for exclusive Higgs production 
at the Large Hadron Collider (LHC) experiments\cite{cox}.
At CDF, the study of diffractive events has been performed by tagging events with either a rapidity gap or a leading hadron.

\section{Diffractive structure functions}

\begin{figure}[tp]
\epsfxsize=1.0\textwidth
\centerline{\epsfig{figure=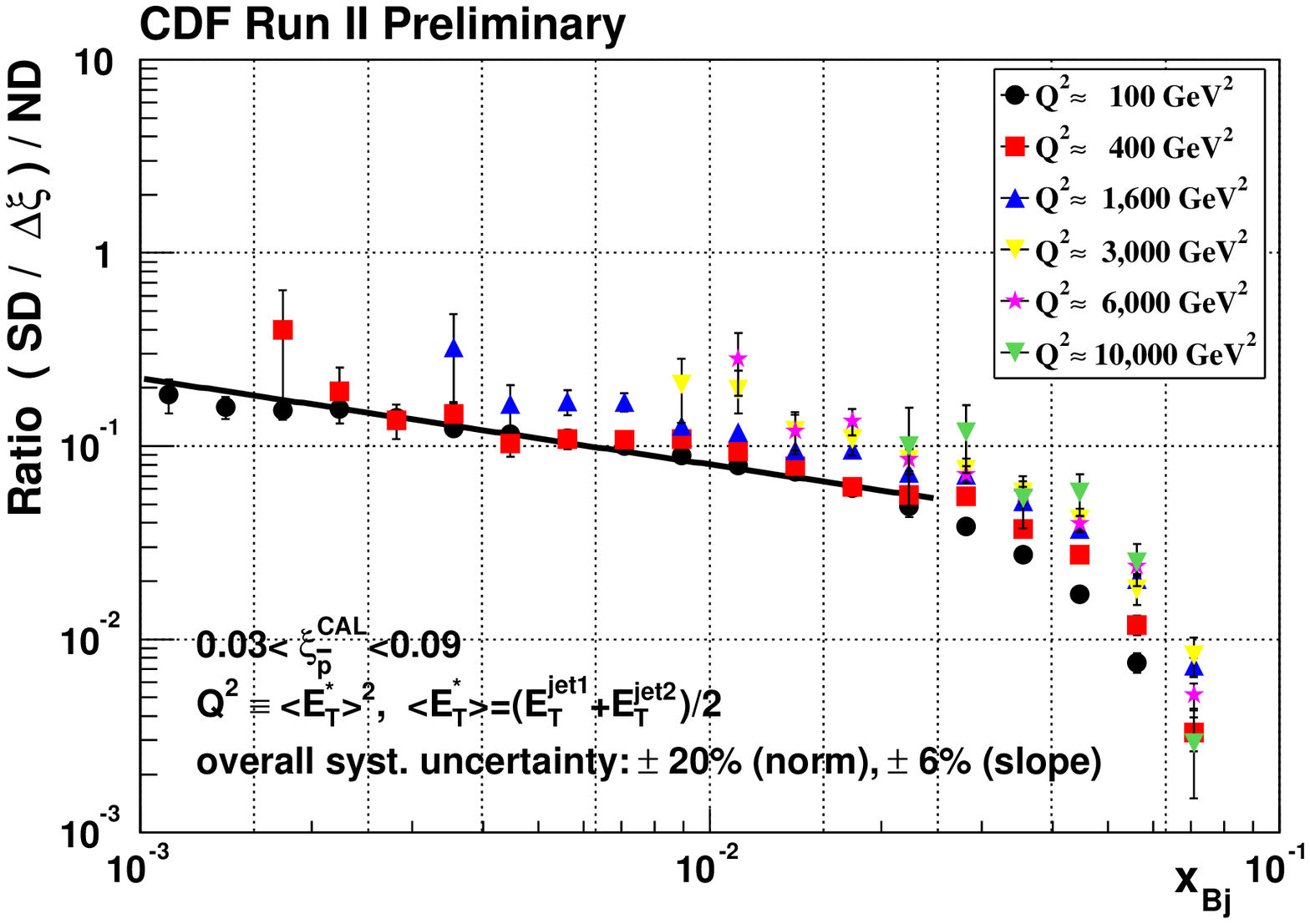,width=0.59\hsize} %\hspace*{0.3cm}
            \epsfig{figure=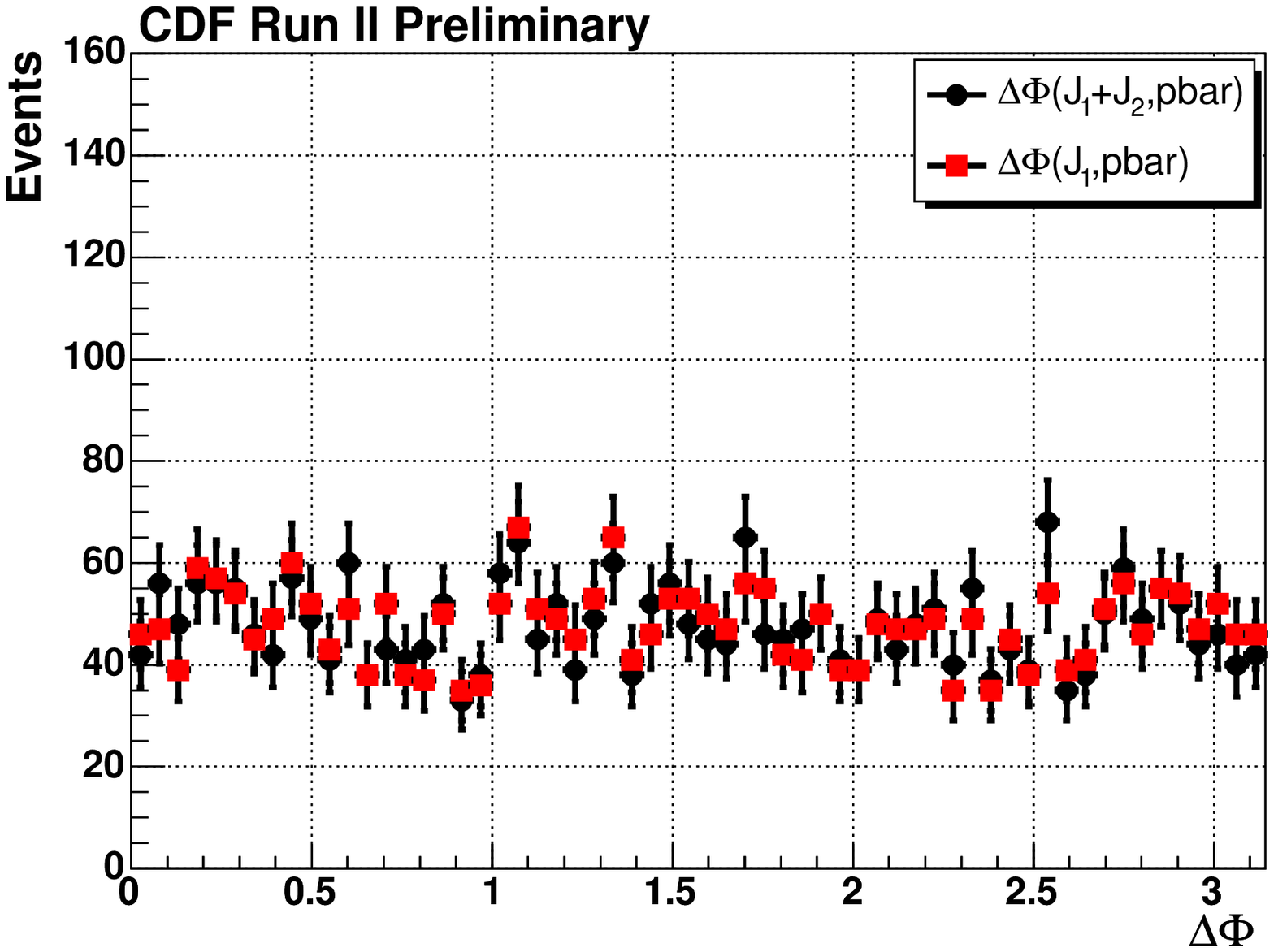,width=0.57\hsize}}
\caption{\label{fig:dsf}
{\em Left}: Ratio of diffractive to non-diffractive dijet event rates
as a function of $x_{Bj}$ (momentum fraction of struck parton in the antiproton) for different values of $E{_T}^2 \equiv Q^2 $;
{\em Right}: Azimuthal angle difference between the jets and the outgoing antiproton in the RP+J5 sample.
The jet angle is that of the leading jet (red squares) or the average of the angles of the two leading jets (black circles).}
\end{figure}

The gluon and quark content of the interacting partons can be investigated by comparing 
single diffractive (SD) and non-diffractive (ND) events.
SD events are triggered on a leading antiproton in the Roman Pot Spectrometer (RPS)\cite{fd} 
and at least one jet, while the ND trigger requires only a jet in the calorimeters.
The ratio of SD to ND dijet production rates ($N_{jj}$) is proportional to the ratio 
of the corresponding structure functions ($F_{jj}$),
$R_{\frac{SD}{ND}}(x, \xi, t)= \frac{N_{jj}^{SD}(x, Q^2, \xi, t)}{N_{jj}(x, Q^2)}
\approx \frac{F_{jj}^{SD}(x, Q^2, \xi, t)}{F_{jj}(x, Q^2)}$,
and can be measured as a function of the Bjorken scaling variable $x\equiv x_{Bj}$\cite{xbj}.
In the ratio, jet energy corrections approximately cancel out, thus avoiding dependence on Monte Carlo (MC) simulation.
Results are consistent with those of Run~I\cite{run1_dsf}, hence confirming a breakdown of factorization. 
In Run~II, the jet $E_T$ spectrum extends to $E_T^{\rm jet}\approx 100$~GeV. 
Preliminary results indicate that the ratio does not strongly depend on $E_T^2\equiv Q^2$ 
in the range $100<Q^2<10,000$~GeV$^2$ (Fig.~\ref{fig:dsf}, left).
The relative normalization uncertainty cancels out in the ratio, and the results indicate that the $Q^2$ evolution, 
mostly sensitive to the gluon density, is similar for the proton and the pomeron.

\section{Measurement of $|t|$ distributions}

\subsection{Dynamic alignment of Roman Pot detectors}
\begin{figure}[tp]
\epsfxsize=1.0\textwidth
\centerline{\epsfig{figure=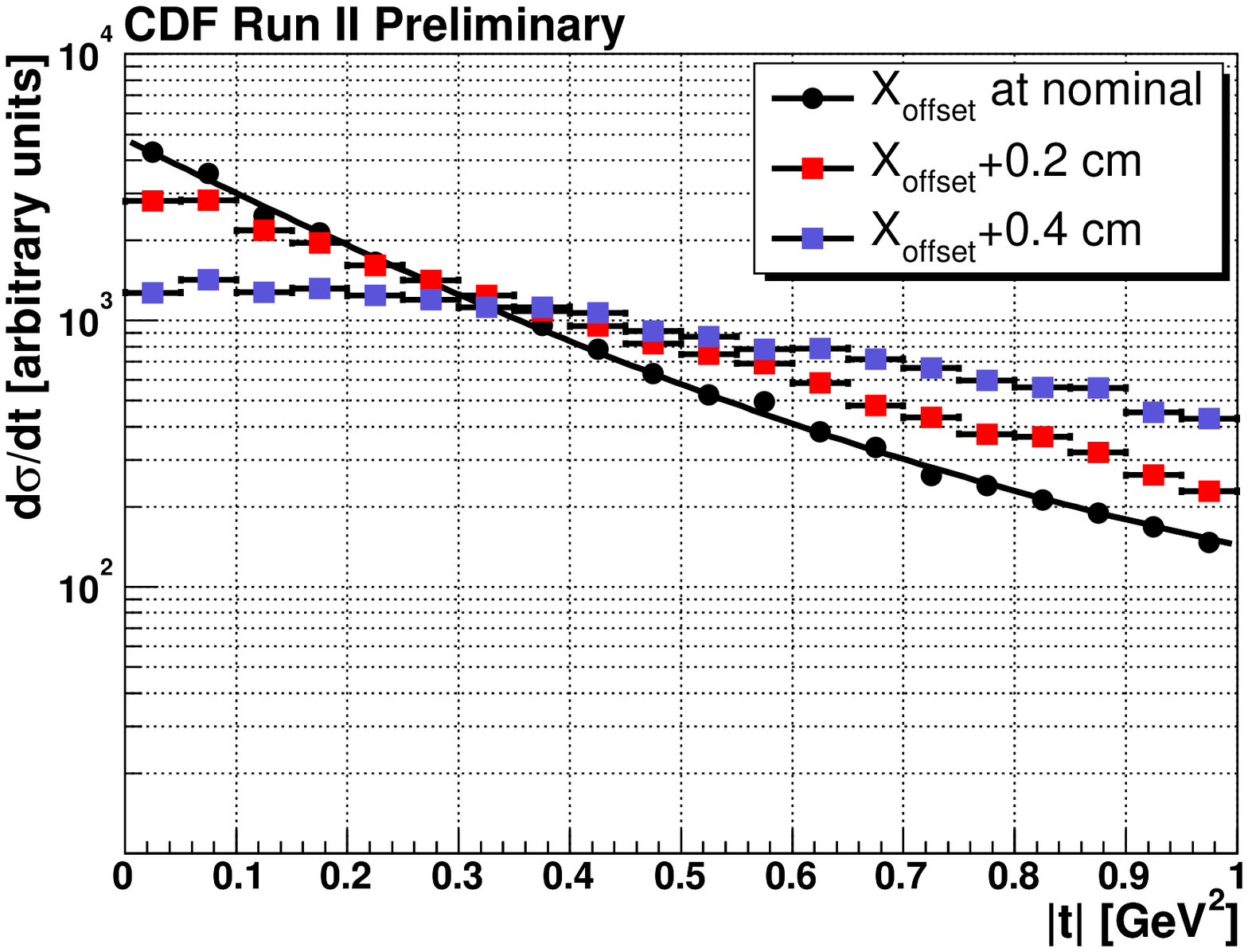,width=0.59\hsize}
            \epsfig{figure=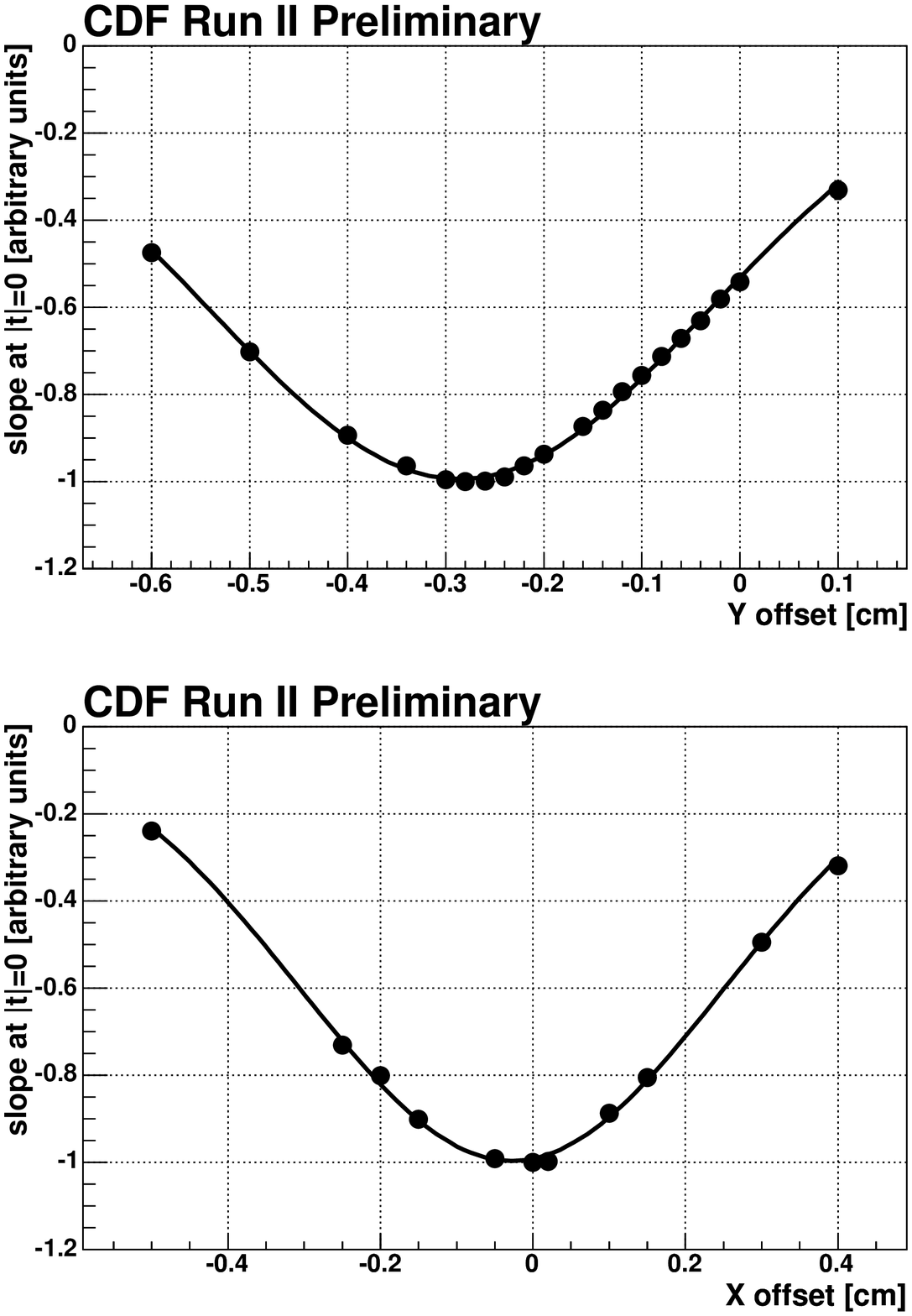,width=0.32\hsize}
}
\caption{\label{fig:talign}
{\em Left}: $t$-distribution of reconstructed RPS tracks for positive $X_{\rm offset}$ shifts;
{\em Right}: $|b|$ slope versus Y $(top)$ and X $(bottom)$ offsets.
}
\end{figure}

The antiproton fractional momentum loss, $\xi$, and four-momentum transfer squared, $t$, of SD events
can be determined from tracks reconstructed in 
the RPS and the position of the
event vertex at the Interaction Point (IP) using the beam transport matrix between the IP and RPS.
Crucial for this measurement is the determination of the detector alignment with respect to the beam.  
The RPS detectors can be aligned by seeking a maximum of the $d\sigma/dt$ distribution at $t=0$ 
for SD events (Fig.~\ref{fig:talign}, left).
Offsets in both the $X$ and $Y$ coordinates of the RPS detectors with respect to the beam are adjusted 
until a maximum for $|d\sigma/dt|$ 
is found at $t=0$, when the RPS fiber tracker is correctly aligned with respect
to the beam (Fig.~\ref{fig:talign}, right). This innovative method is very precise and quite general, and can be used to 
accurately calibrate the RPS detector position with respect to the beam in CDF using current data
or in future experiments at the LHC. 
The accuracy of the RPS alignment calibration 
%can be infinitely precise and depends on the detector resolution.
at CDF is $\Delta \rm X\approx \pm 30~\mu$m and $\Delta \rm Y\approx \pm 30~\mu$m, respectively.

\subsection{$|t|$ distributions}
\begin{figure}[tp]
\epsfxsize=1.0\textwidth
\centerline{\epsfig{figure=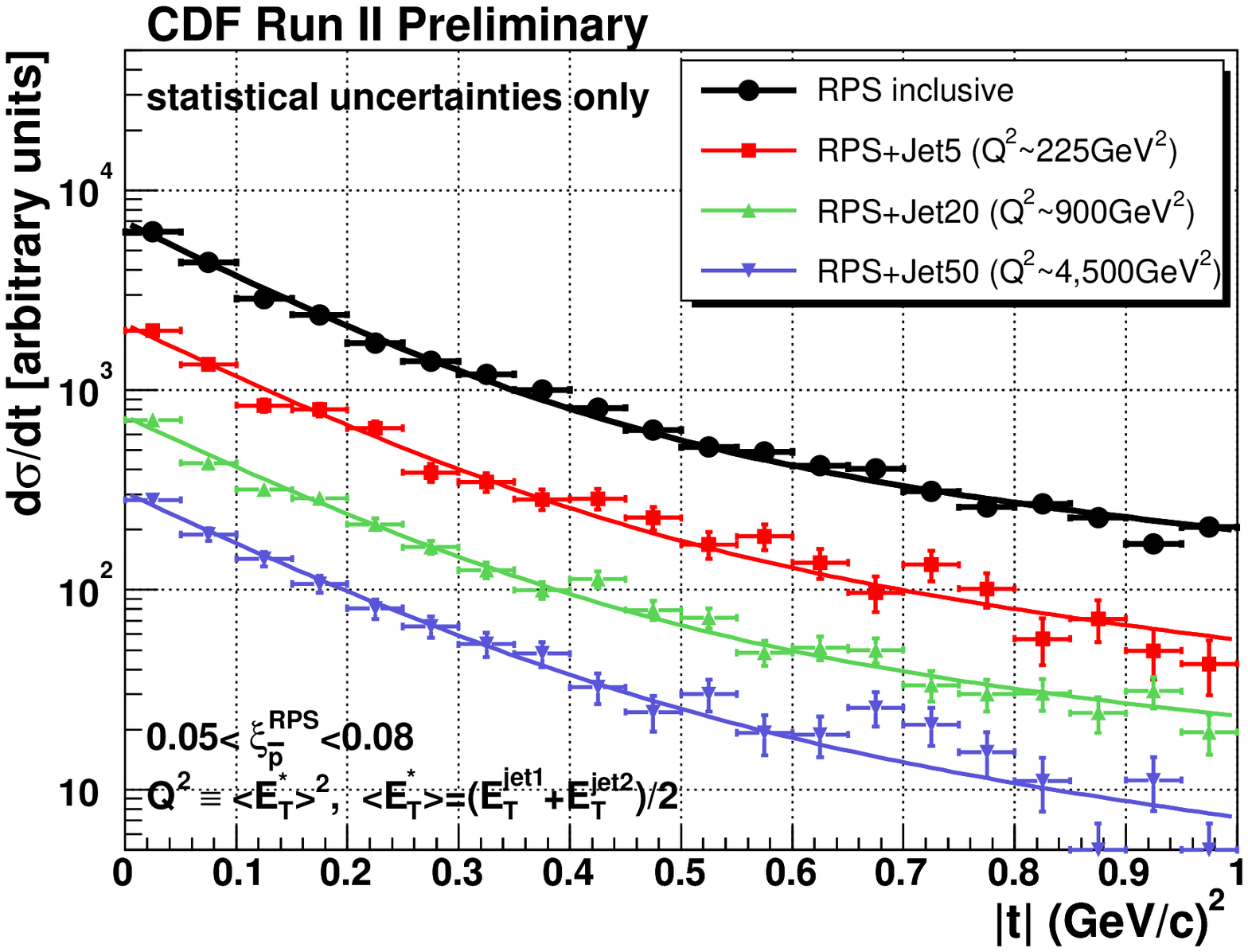,width=0.59\hsize} %\hspace*{0.3cm}
            \epsfig{figure=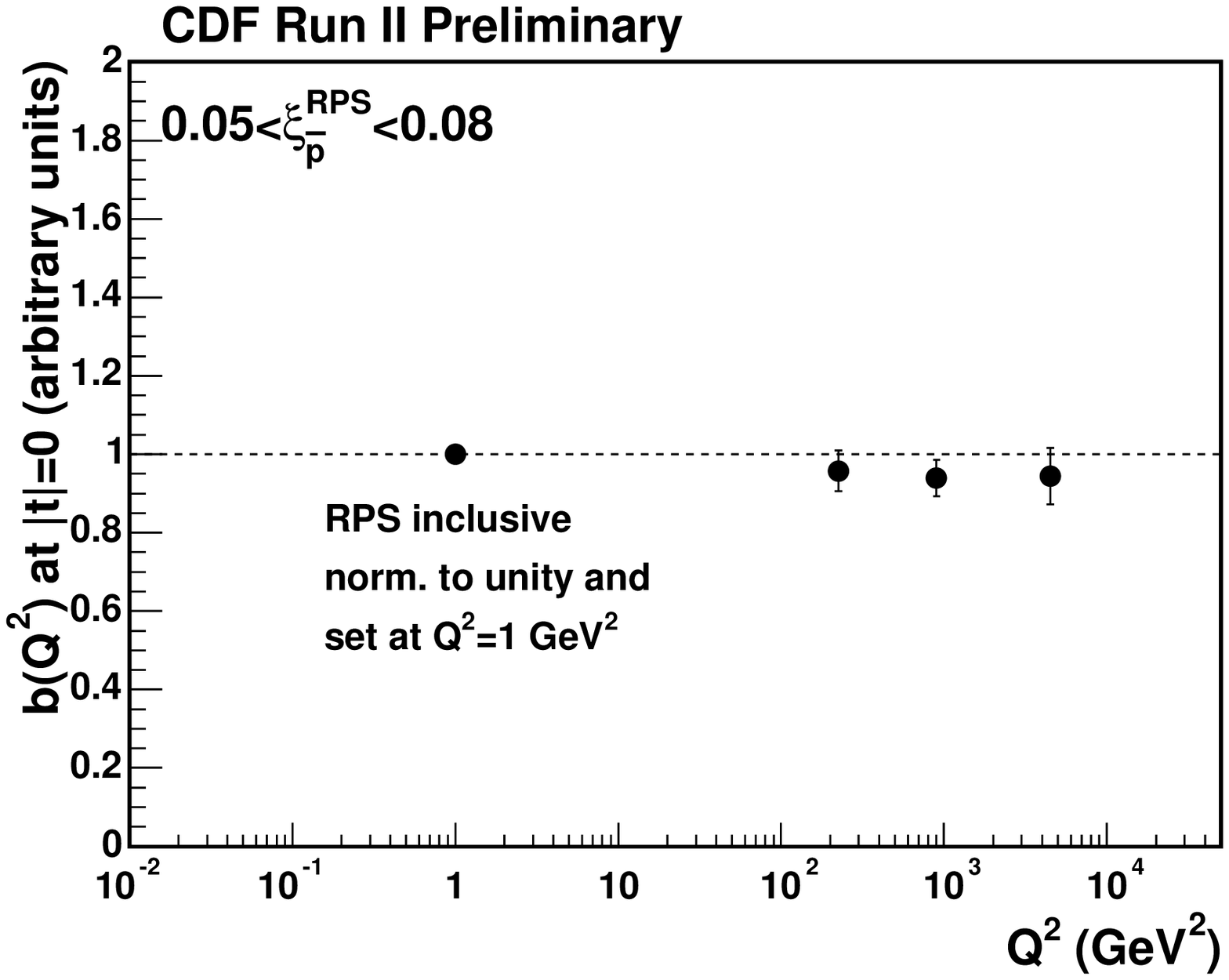,width=0.59\hsize}}
\caption{\label{fig:tdist}
{\em Left}: $|t|$-distribution measurement for soft and hard SD events;
{\em Right}: $b$-slope at different $Q^2$ values %at $|t|=0$~GeV$^2$ 
(slope of RPS inclusive data is normalized to $b=1$).
}
\end{figure}

SD events studied contain both soft and hard diffractive interactions. 
An exponential fit with a slope $b$ and arbitrary normalization agrees well with the data in the region $0<|t|<1$~GeV$^2$ 
for different data samples in which the mean dijet transverse energy is increasingly larger (Fig.~\ref{fig:tdist}, left).
The measured $|t|$ distribution does not show diffractive minima or ``dips'',
which could have been caused by the interference terms of imaginary and real parts of the interacting partons.
When comparing soft and hard diffractive events, results show that the $b$ parameter 
is equal within uncertainties up to $Q^2\approx 4,500$~GeV$^2$ (Fig.~\ref{fig:tdist}, right).

The azimuthal angle difference, $\Delta\Phi$, between the jets and the outgoing antiproton 
is a flat distribution shown in Figure~\ref{fig:dsf} (right), and it does not show any correlation.

\section{Exclusive dijet production}

\begin{figure}[tp]
\epsfxsize=1.0\textwidth
\centerline{\epsfig{figure=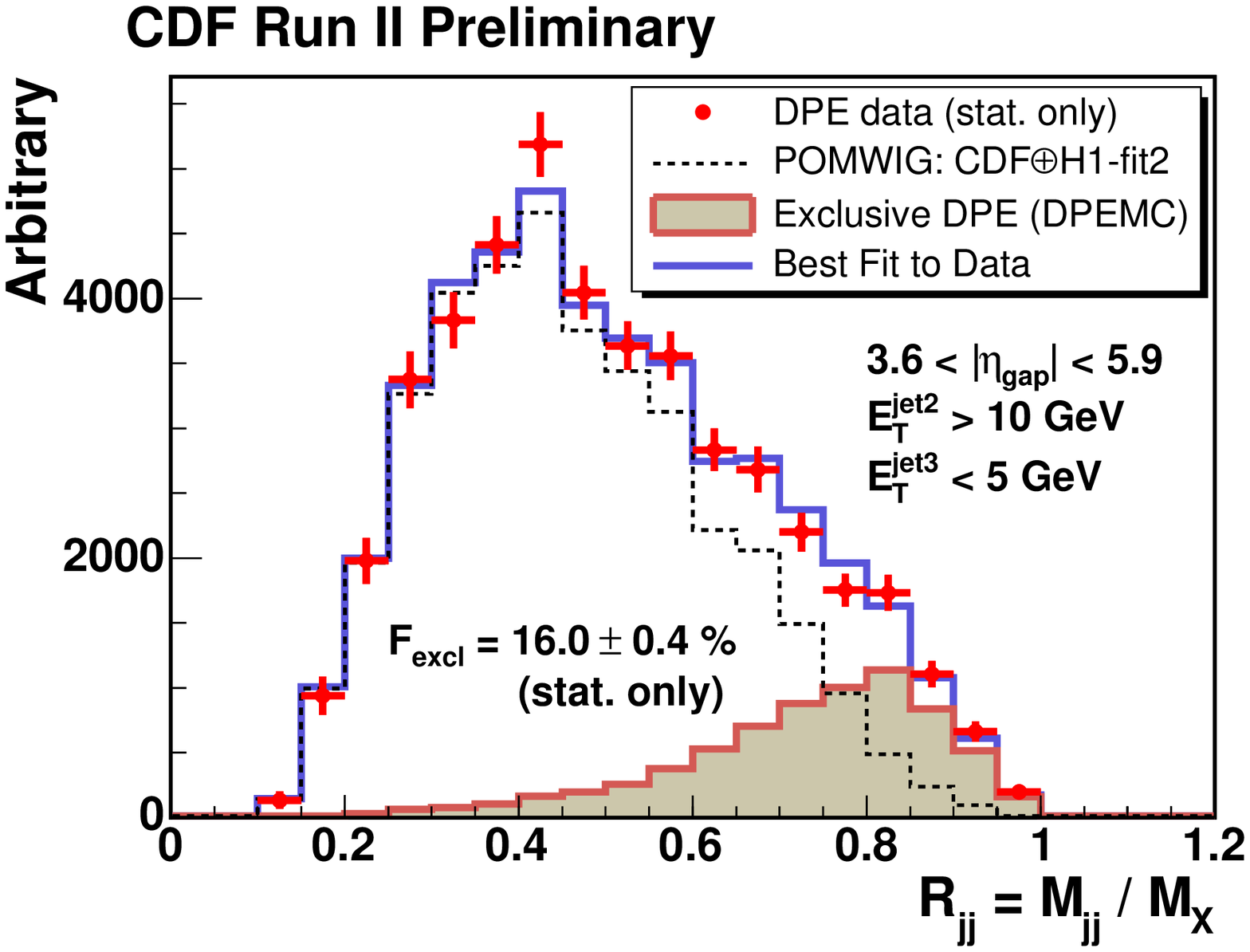,width=0.39\hsize}
	    \epsfig{figure=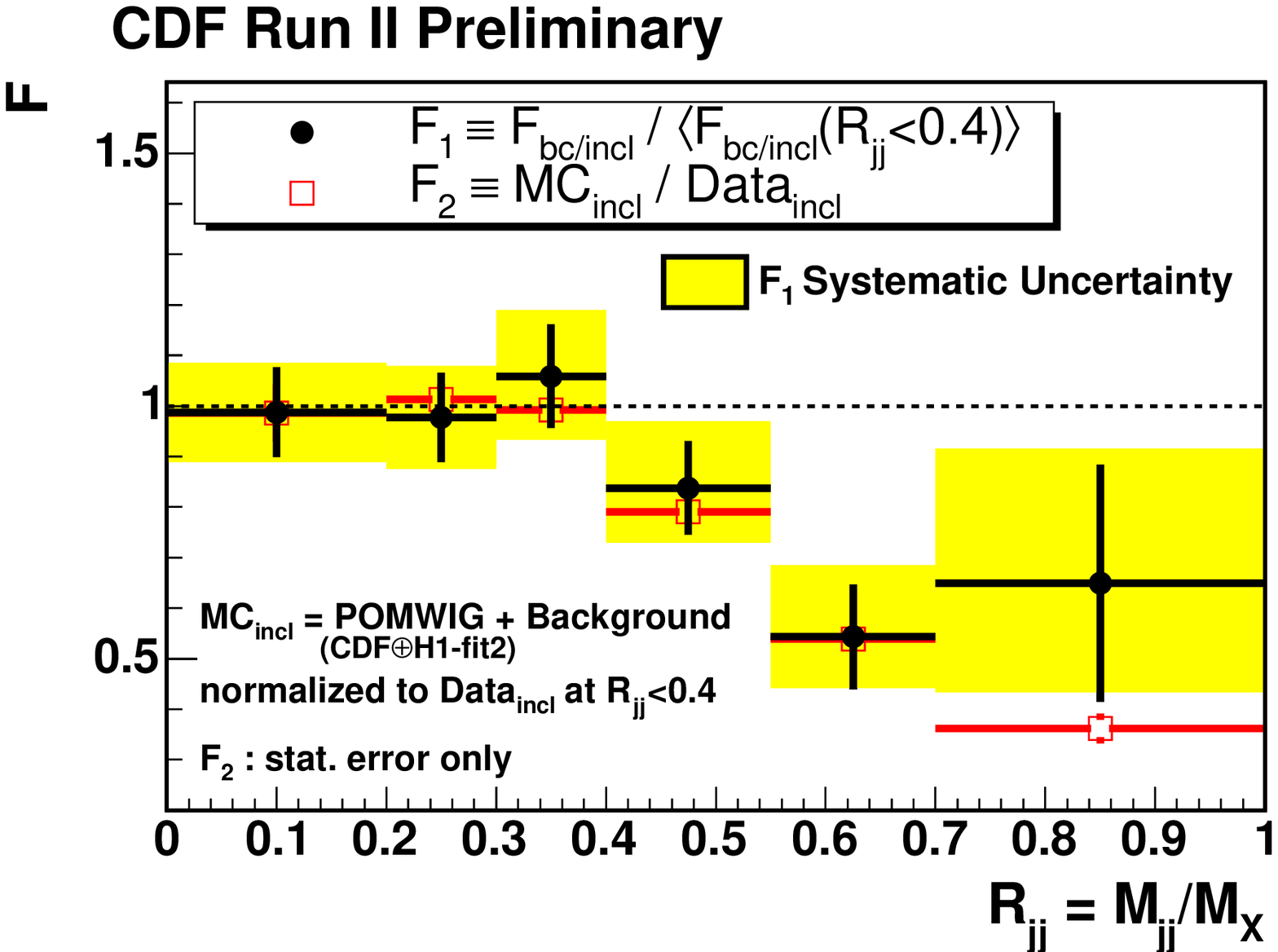,width=0.39\hsize}
	    \epsfig{figure=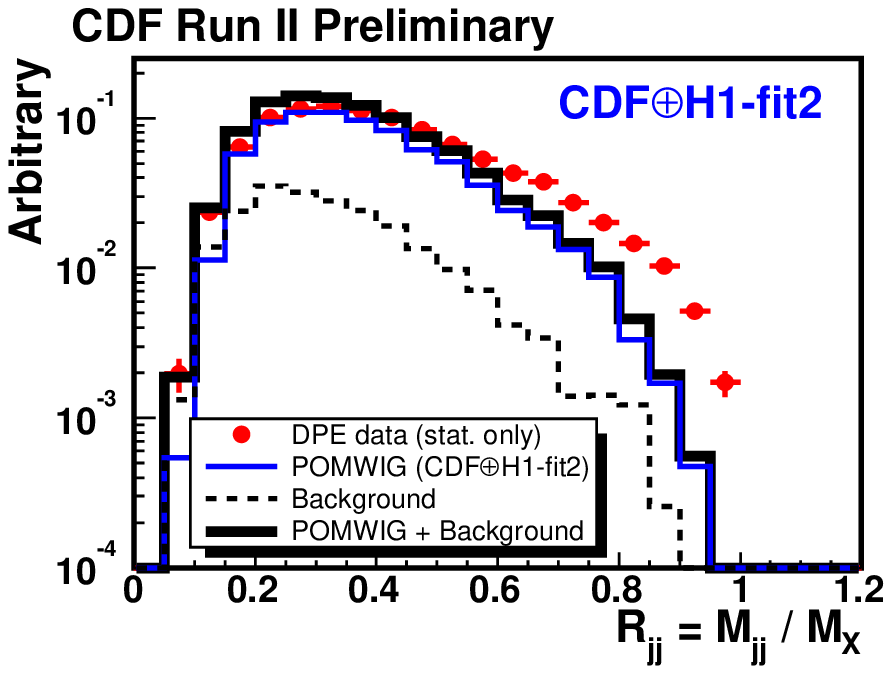,width=0.39\hsize}}
\caption{\label{fig:excl}
{\em Left}: dijet mass fraction in DPE data (points) and best fit (solid) obtained from POMWIG MC events (dashed) 
and exclusive dijet MC events (shaded);
{\em Center}: normalized ratio of heavy flavor jets to all jets as a function of dijet mass fraction.
{\em Right}: $R_{jj}$ distribution for the data (points) and POMWIG MC prediction (thick histogram), 
composed of DPE dijet events (thin) and non-DPE events (dashed). Data and MC are normalized to the same area.
}
\end{figure}

Exclusive production at the Tevatron can be used as a benchmark to establish predictions on exclusive diffractive Higgs production, 
a process with a much smaller cross section\cite{kmr}.
This is the case in Higgs production through the Double Pomeron Exchange (DPE) processes
$p\overline{p} \rightarrow p H \overline{p}$ (or $p p \rightarrow p H p$),
where the leading hadrons in the final state are produced at small angles with respect to the direction of the incoming particles
and two large forward rapidity gap regions are present on opposite sides of the interaction.
The Higgs production process through $g g\rightarrow H$ is replaced by the 
$gg\rightarrow {\rm jet~ jet}$ process, with a much larger production cross section.
The characteristic signature of this type of events is a leading nucleon and/or a rapidity gap on both 
forward regions, and it results in an exclusive dijet final state produced together
with both the leading proton and anti-proton surviving the interaction and escaping in the very forward region.
The CDF RPS spectrometer can tag the anti-proton, while
the proton is inferred by the presence of an adjacent large ($\Delta\eta>3$) rapidity gap.
The dijet mass fraction ($R_{jj}$), defined as the dijet invariant mass ($M_{jj}$) divided by the mass of the entire system,
$M_X =\sqrt{\xi_{\overline{p}}\cdot\xi_p\cdot s}$, is calculated using all available energy in the calorimeter.
If jets are produced exclusively, $R_{jj}$ should be equal to one. Owing to hadronization effects, 
underlying event energy spilling out of the jet reconstruction cone, 
and radiation from the jets, the sharp peak from exclusive production is smeared out to a wider distribution.
The search is performed by comparing data with MC expectations.
At large $R_{jj}$ values, the excess of events in the data with respect to inclusive DPE dijet production, 
which is described by POMWIG\cite{pomwig} MC,
is well accounted for by the DPEMC\cite{dpemc} (or equivalently ExHuME\cite{exhume}) 
MC sample of exclusive events (Fig.~\ref{fig:excl}, left).
%The third jet veto is used because the exclusive MC only generates LO $gg\rightarrow gg$ process.

The quark/gluon composition of dijet final states can be used to provide additional information on exclusive dijet production. 
At leading order (LO) $gg\rightarrow gg$ process is dominant while $gg\rightarrow q\overline{q}$ is strongly suppressed. 
This ``suppression'' mechanism can be used to improve the sensitivity to exclusive production.
Thanks to high tagging efficiency of heavy flavor jets and low mistag rate, $b/c$-quarks are selected.
The ratio ($F_{\rm bc/incl}$) of heavy flavor tagged jets divided by all inclusive jet events is measured as a function of $R_{jj}$
and is normalized to the weighted average in the region $R_{jj}<0.4$.
In the large mass fraction region ($R_{jj}>0.6$) a significant ``dip'' is observed in the data, 
indicating a contribution due to exclusive production 
(Fig.~\ref{fig:excl}, center).
The result is compared with the ratio of the inclusive dijets, where $F_2$ is the ratio of the inclusive MC events to the data 
(Fig.~\ref{fig:excl}, right).

\section{Exclusive photon pair production}

Another process which can be used as ``standard candle'' is 
exclusive diphoton events, $p\overline{p}\rightarrow p\gamma\gamma\overline{p}$.
The final state is cleaner than in exclusive dijet production as hadronization effects are absent, 
but the expected cross section is smaller.
CDF has performed a search in this channel
by requiring nothing else\footnote{In this search, the leading hadrons are not detected.}
except two electromagnetic (EM) calorimeter towers above threshold in the final selection. 
Three exclusive $\gamma\gamma$ candidate events with $E_T>5$~GeV are found with 
no tracks pointing at the clusters, with a small expected background.
The purely QED $p\overline{p}\rightarrow p e^+ e^- \overline{p}$ process is mediated 
through $\gamma\gamma\rightarrow e^+ e^-$ scattering and constitutes a good control sample: 
16 exclusive $e^+e^-$ candidate events are selected in the data with a small background of $2.1^{+0.7}_{-0.3}$ events.
The cross sections measured,
$\sigma (\gamma\gamma)=0.14^{+0.14}_{-0.04} {\rm (stat)} \pm 0.03 {\rm (syst)}$~pb and
$\sigma (e^+ e^-)= 1.6^{+0.5}_{-0.3} {\rm (stat)} \pm 0.3 {\rm (syst)}$~pb, 
are in agreement with expectations from exclusive ExHuMe and QED LPAIR MCs, respectively.


\begin{thebibliography}{0}
\bibitem{xbj}
For each event, $x_{Bj}$ is evaluated from the $E_T$ and $\eta$ of the jets using the equation
$x_{Bj}=\frac{1}{\sqrt{s}}\sum_{i=1}^3E_T^ie^{-\eta^i}$.
\end{thebibliography}
\end{document}